\documentclass[pra,aps,twocolumn,superscriptaddress,showpacs]{revtex4}
\usepackage{graphicx}
\usepackage{amsmath}

\newcommand{\eq}{\begin{equation}}
\newcommand{\fine}{\end{equation}}

\begin{document}

\title{
\vspace{2cm
\begin{flushright}
CERN-PH-TH/2017-215
\end{flushright}
\vspace{3cm}
\bf \LARGE A Lower Bound on Inelasticity in Pion-Pion Scattering }
\vspace{.2cm}}
\date{5 Oct. 2017}
\author{Andr\'e Martin }
\email{martina@mail.cern.ch} \affiliation{Theoretical Physics Division,CERN, Geneva}

\author{S. M. Roy}
\email{smroy@hbcse.tifr.res.in} \affiliation{HBCSE,Tata Institute of Fundamental Research, Mumbai}

\begin{abstract}

Assuming that the pion-pion scattering amplitude and its absorptive part are analytic inside  an ellipse in $t$- plane 
with foci $t=0$, $u=0$ and right extremity $t=4 m_{\pi}^2 +\epsilon $, ($\epsilon > 0$), except for cuts prescribed by Mandelstam 
representation for $t\geq 4 m_{\pi}^2$, 
$u\geq 4 m_{\pi}^2$ , and bounded by $s^N$ on the boundary of this domain, we prove that for $s\rightarrow \infty$,
\begin{equation}
 \sigma_{inel} (s) > \frac{Const}{s^{5/2} }\exp {[-\frac{\sqrt{s}}{4} (N+5/2) \ln {s} ]} .\nonumber
\end{equation}

 - - - - - - - - - - - - - - - - - - - - - - - - -  {\it Dedicated to the memory of Stanley Mandelstam.}
                                      
\end{abstract}


\maketitle

{\bf I. Introduction}.

\begin{figure*}[!]
\includegraphics[width=2.0\columnwidth]{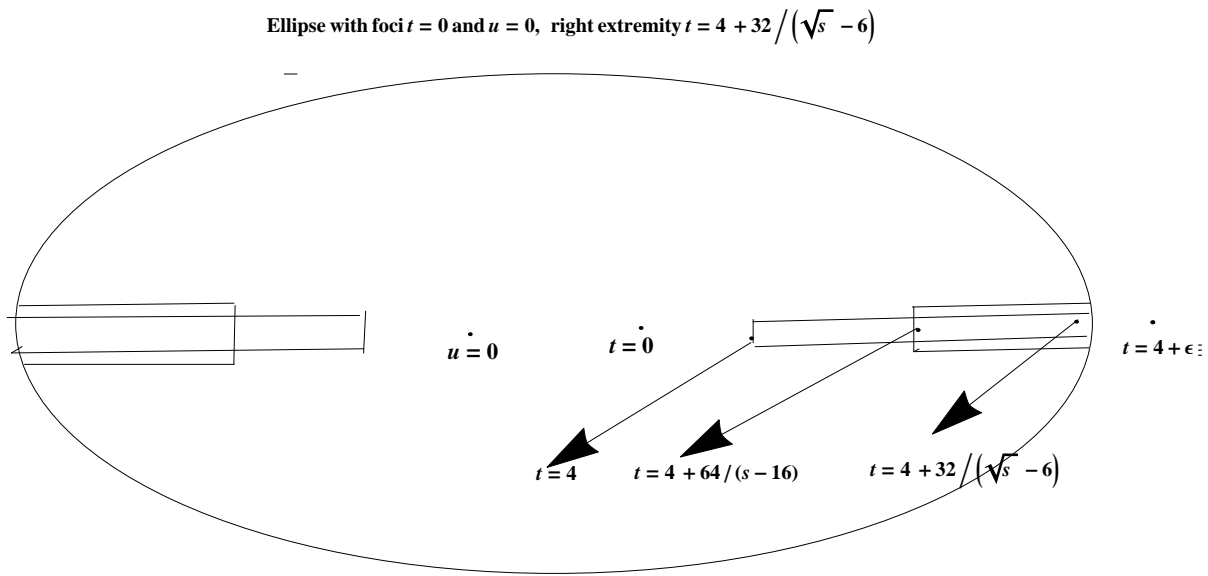}
\caption{The amplitude $F(s,t)$ is assumed to be analytic in $t$ within the ellipse shown except for 
cuts $t\geq 4, u\geq 4$ ; 
its absorptive part $F_s(s,t)$,for $s\geq 20$ is assumed to be analytic in $t$ within the same ellipse 
except for cuts for $t\geq4+\frac{64}{s-16},u\geq4+\frac{64}{s-16} $.The truncated Froissart-Gribov 
formulae for  $f_l(s)$ ,Eq.(\ref{Re f_l}) and $Im f_l(s)$, Eq. (\ref{Im f_l}) follow from this. 
Note that the horizontal and vertical scales in this figure are not the same.
}
\label{Contours}
\end{figure*}

It is well known that if there is no inelasticity, the scattering amplitude must be 
zero.However, there is no quantitative estimate of the amount of inelasticity required.
This is what we try to do.There are various proofs of the fact that the 
scattering amplitude must be zero if there is no inelasticity. A very appealing attempt 
has been made by Cheung and Toll \cite{Cheung}.Their idea is to use repeatedly elastic unitarity 
at all energies to the point where they get an absurd analyticity domain much too large.However,
even after the enlargement of the pion-pion analyticity domain by one of us in 1966 
\cite{Martin1966}, it is not obvious that they have really succeeded.Alexander Dragt 
\cite{Dragt} has a proof which is nice but not quite complete: it uses the fact that 
partial wave amplitudes for very large angular momenta are dominated by the nearest 
singularities in the crossed channel. He needs more analyticity than what has been 
proved from field theory \cite{Martin1966}. For instance, the Mandelstam representation 
\cite{Mandelstam} with a finite number of subtractions is largely sufficient. In fact 
he needs much less than that. Since we shall also use the dominance of the nearest 
singularities for large angular momenta, we state at the same time the assumption
he needs and our assumption. If we use the standard mandelstam variables $s,t,u$ , 
and choose units such that the pion mass $m_\pi =1$, we need fixed energy 
analyticity in an ellipse with foci at $t=0$ and $u=0$ and right extremity at 
$t=4+\epsilon$, minus the obvious cuts  $t\ge 4,u\ge 4$ for the amplitude , and 
$t\ge 4+64/(s-16), u\ge 4+64/(s-16)$ for the absorptive part (see Fig. \ref{Contours}). 
From field theory we only get ,for the 
absorptive part, an ellipse with right extremity at $t=4$ exactly, and for the 
amplitude a region containing $|t|<4$ . In fact for $|t|<4$ fixed-$t$ dispersion 
relations are valid, and with our assumptions they are valid for $|t|< 4 +\epsilon$.
With these assumptions we can prove that there must be inelasticity at energies such 
that $ s> 16 + 64/\epsilon $. For instance, if $\epsilon=12$ (corresponding to the 
full $t$-channel elastic strip), we must have inelasticity for $s>22$.

For simplicity, we look first at 
$ \pi^0 \pi^0$ scattering amplitude $F(s,t)$  where $\pi^0$ is a fictitious iso-spin zero 
neutral pseudoscalar particle. It has the  partial wave expansion,
\begin{eqnarray}
 F(s,t)&=& \sum_{l=0}^{\infty} (2l+1) f_l (s)P_l ( 1+\frac{2t}{s-4})\>; \nonumber \\
 f_l (s)&=& a_l (s)/\rho (s); \>\rho (s)\equiv \frac{2k}{ \sqrt s }=\sqrt{\frac{s-4}{s} }
 \end{eqnarray}
 with the unitarity constraint 
 \begin{eqnarray}
 Im  a_l (s) = | a_l (s) |^2, 4\le s \le 16\>; \nonumber \\
 Im  a_l (s) \ge | a_l (s) |^2, s \ge 16 \>.
 \end{eqnarray}
 The optical theorem gives,
 \begin{equation}\label{optical}
  \sigma _{tot}=\frac{8\pi}{k^2}\sum_{l=0}^{\infty} (2l+1) Im a_l (s)=\frac{16\pi}{k\sqrt s}F_s (s,0),
 \end{equation}
where $F_s(s,t)$ denotes the $s$-channel absorptive part $Im F(s,t)$.
 Similar unitarity conditions hold in the $t$ and $u$ channels.
   The normalization specified by the above choice of $\rho (s)$ corresponds to 
$F (4,0)=S $-wave scattering length $a_0$. For the generalization to real pions of iso-spin 1, 
we shall use the same normalizations as above, with $F(s,t), f_l(s),a_l(s),\sigma_{tot},
A_s(s,0),F(4,0),a_0$ being replaced respectively by the corresponding quantities with 
superscript $I$, e.g. $F^I (s,t),.,a_0^I$.

Our strategy will be the following.we write the partial-wave amplitudes as well as their 
imaginary parts as contour integrals along the ellipse mentioned above, and add the contribution of 
the cuts (see fig. 1).Then we try to get an upper bound on the partial wave amplitude $f_l$ for 
which we need an upper bound $B(s)$ on the whole ellipse. We also seek 
 a lower bound on it's imaginary part  $Im f_l$ , for which we need a bound on the 
 discontinuity of the absorptive part which is nothing but the Mandelstam double spectral function.
 In fact this is what is missing in the work of Dragt \cite{Dragt}.This will be done in the next section.

{\bf Domain of Positivity of the Double Spectral Function and a Lower Bound} 
In the first part, we recall the results of Mahoux and one of us \cite{Mahoux-Martin} on 
the domain of positivity of the double spectral function.
 For $s>20 $, the absorptive part in the $s-$channel 
has a cut beginning at 
\begin{equation}
 t=4+ \frac{64}{s-16} .
\end{equation}
From $t=4$ to $t= 4+\epsilon <16$, the discontinuity across the cut is given by  
 the Mandelstam form of the $t$-channel elastic unitarity condition 
 on one of the double spectral functions $ \rho _{st} (s,t) $,

 \begin{equation} \label{Mandelstam}
    \rho _{st}  (s,t) = \frac{2 \rho (t)}{\pi} \int \int \frac { dz_{1} dz_{2} } 
   {\sqrt {H(z,z_{1},z_{2})} } F_s (s_1,t)F_s (s_2,t)^*,
\end{equation}
where ,
\begin{eqnarray}
&\rho (t)=\sqrt{\frac{t-4}{t} } ,z=1+\frac{2s}{t-4},z_0\equiv 1+\frac{8}{t-4},\nonumber \\
&\>z_{i}=1+(2s_i)/(t-4),i=1,2 \>
\end{eqnarray}
and
\begin{eqnarray}\label{H}
&H(z,z_{1},z_{2})=z^2+z_{1}^{2} + z_{2}^{2}-1-2 z z_{1}z_{2} \nonumber\\
& =(z-z_{+})(z-z_{-}) .
\end{eqnarray}
with ,
\begin{equation}\label{z+-}
 z_{\pm}=z_{1}z_{2} \pm \sqrt { ( z_1 ^{2}-1 ) ( z_{2} ^{2}-1 )  }. 
\end{equation}
The domain of integration in the $z_1,z_2$ plane is bounded by the three 
 lines,
\begin{equation}\label{zdomain}
 z_1>z_0,\>z_2>z_0,\>,z>z_+ \>.
\end{equation}
If we define 
\begin{equation}
 z=\cosh \theta ; \> z_i=\cosh \theta _i,\> i=0,1,2,
 \end{equation}
 then, the region (\ref{zdomain} ) becomes just a triangle in 
 the $\theta_1,\theta_2$ plane bounded by the lines,(see Fig.\ref{Regions A,B,C})  
 \begin{equation} \label{thetadomain}
  \theta_0 \le \theta_1,\theta_0 \le \theta_2,\>\theta _1 +\theta _2 \le \theta \>.   
 \end{equation}
These inequalities imply that for $i=1,2$, $  \theta_0 \le \theta_i \le \theta -\theta_0 $,i.e.
\begin{equation} 
\label{domain}
 z_0 \le z_i \le zz_0 - \sqrt { ( z ^{2}-1 ) ( z_{0} ^{2}-1 )  }.
\end{equation}
They also imply that $\theta \ge 2 \theta _0 $ which gives the 
boundary curve of the spectral region
\begin{equation}
 s \ge \frac {16 t}{t-4}.
\end{equation}

 It will be crucial to recall the observation of Mahoux and Martin 
 \cite{Mahoux-Martin} that when  $\theta \le 3\theta _0$ , the inequalities (\ref{thetadomain} ) imply 
 that only values of $\theta_i \leq 2 \theta_0 $ for $i=1,2$ , i.e. only values of $F_s (s_i,t) $ 
 outside the spectral region for $i=1,2$ are needed to compute the double spectral function. In this 
 region, the convergent partial wave expansion,
 \begin{equation}
   F_s(s_i,t)= \sum_{l=0}^{\infty} (2l+1) Im f_l (s_i)P_l ( 1+\frac{2t}{s_i-4}), \>i=1,2 ,
 \end{equation}
 the positivity of $ Im f_l (s_i)$  and the inequalities $P_l ( 1+2t/(s_i-4)) >1 $ imply that 
 $F_s(s_i,t) > 0 $ for $i=1,2$. Hence,the double spectral function $ \rho _{st}  (s,t)$ 
 is positive when  $\theta \le 3\theta _0$, i.e. for
 \begin{eqnarray}\label{domain1}
 && 4\le t \le 16,\>and \>\frac {16 t}{t-4} \le s \le 4 \big (\frac{3t+4}{t-4} \big )^2 ,\nonumber\\
 && i.e. 4 +64/(s-16) \le t \le 4+32/(\sqrt{s} -6) .
 \end{eqnarray}
 
 Since $ \rho  (s,t)$ is symmetrical in its arguments, it is also 
positive for,
\begin{equation}
 4\le s \le 16,\>\frac {16 s}{s-4} \le t \le 4 \big (\frac{3s+4}{s-4} \big )^2 .
 \end{equation}

 {\bf II.  Lower bound on inelasticity }.
 
 We shall now obtain a lower bound on $\rho (s,t)$ in the domain  (\ref{domain1}) in terms of total cross sections $ \sigma_{tot} (s_1) ,\sigma_{tot} (s_2)$, 
 where $s_1, s_2$ are such that Eqn. (\ref{domain}) holds for the corresponding $z_1, z_2$. We then deduce 
 a lower bound on inelasticity . It will then follow that if there is no inelasticity at one (and only one) energy in the $s$-channel 
 ($s>20$), the double spectral function must vanish in the range $t=4 +64/(s-16)$ to $ t=4+32/(\sqrt{s} -6) $ , and hence that there is 
 an interval of energy given by (\ref{domain}) in which the total cross section vanishes. This is impossible and hence the scattering 
 amplitude is zero.It must be realized that only a small fraction of Mandelstam representation is used.
 
 Now, the question which was asked to one of us (AM) by  Miguel F. Paulos, during a conference organized 
 by Jo$\tilde{a}$o Penedones at EPFL , Lausanne was, if the inelastic cross section could be 
 arbitrarily small. We want to show that, with some assumptions much weaker 
 than the Mandelstam representation, but slightly stronger than what has been 
 proved from local field theory, there exists a lower bound to inelasticity ,
 
 \begin{equation}
  \sigma_{inelastic} > C \exp {(-\sqrt{(s/s_0)} \log (s/s_0) )} .
 \end{equation}

  The strategy we shall use is based on the  results of Mahoux and Martin \cite{Mahoux-Martin} 
  on positivity of double spectral functions, and on the research made by Dragt \cite {Dragt}, 
  viz. that the real and imaginary parts of the 
partial wave amplitudes are dominated by the  contributions of the nearby cuts 
in the crossed channel:

\begin{eqnarray}
 from \> t=4\> to\> t=t_M(s)\>for Re f_l \>and\>  f_l \nonumber \\
 from \> t=4+\frac{64}{s-16}\> to\> t=t_M(s)\>for Im f_l ,
\end{eqnarray}
where,
\begin{equation}
 t_M(s)\equiv 4+\frac{32}{\sqrt{s}-6}.
\end{equation}

  {\bf Estimates of $f_l(s)$ and $Im f_l(s)$}
  We shall use a truncated Froissart-Gribov representation for $Re f_l(s)$ 
  and $Im f_l(s)$. It follows from analyticity of $F(s,t)$ in $t$ within an 
  ellipse with right extremity $t=t_M (s) $ and foci $t=0$ and $u=0$, except for cuts 
  $4 \leq t \leq t_M (s)$ and  $4 \leq u \leq t_M (s)$ .For $l$ even, 
  \begin{eqnarray}\label{Re f_l}
   f_l(s)=\frac{1}{\pi k^2} \int _4 ^{4+\frac{32}{\sqrt{s}-6}} 
   Q_l(1+\frac{2t}{s-4}) F_t(s,t) dt \nonumber \\
   +\frac{1}{4 i\pi k^2} \int _\Gamma 
   Q_l(1+\frac{2t}{s-4}) F(s,t) dt .
  \end{eqnarray}
  
where $\Gamma$ is an ellipse with foci at $t=0$ and $u=0$, and right 
extremity at $t=4+\frac{32}{\sqrt{s}-6}$ (see figure \ref{Contours}) .

Hence,

 \begin{eqnarray}\label{Im f_l}
   Im f_l(s)=\frac{1}{\pi k^2} \int _{4+\frac{64}{s-16}} ^{4+\frac{32}{\sqrt{s}-6}} 
   Q_l(1+\frac{2t}{s-4}) \rho (s,t) dt \nonumber \\
   +\frac{1}{4i\pi k^2} \int _\Gamma 
   Q_l(1+\frac{2t}{s-4}) F_s (s,t) dt .
  \end{eqnarray}
 where $\rho (s,t)$ is given by the Mandelstam equation (\ref{Mandelstam}).  
As noted earlier, if $s$  is in the Mahoux-Martin domain (\ref{domain1}), 
$\rho (s,t)$ is positive. 

Now we postulate that $F(s,t)$ and $F_s(s,t)$ are bounded by $B(s)$ in the 
ellipse $\Gamma$.The behaviour of  $B(s)$ for $s\rightarrow \infty$ will be 
discussed later.Now we need some estimates on the $Q_l$'s. We prove that,
for $z$ real and $>$ 1, (see Appendix)
\begin{eqnarray}
 \sqrt{\frac{\pi }{ 2l+2} } \frac{1}{(z+\sqrt{z^2-1} )^{l+1} } < Q_l(z) < 
 \nonumber \\
 \frac{1}{(z+\sqrt{z^2-1} )^{l} } \frac{1}{2}|\ln |\frac{z+1 } {z-1 } | | .
\end{eqnarray}
and for $z= \cosh ( (\theta _1 + i \theta _2 )) $, (see Appendix),
\begin{equation}
| Q_l (  \cosh ( (\theta _1 + i \theta _2 ))) |< | Q_l (  \cosh ( (\theta _1 )) )|.
\end{equation}
This means that on an ellipse with foci $ cos \theta =\pm 1$ the modulus of 
$ Q_lcos \theta )$ is maximum at the right extremity.
 
We can get a bound on $|f_l|$

\begin{equation}
 |f_l|< \frac{1}{4 \pi k^2} Q_l (1+\frac{8}{s-4} )B(s) L(s)
\end{equation}
where $L(s)$ is the perimeter of the ellipse with extremities at
\begin{equation}
 cos \theta _s =\pm \big (1+ \frac{1}{2k^2}(4+\frac{32}{\sqrt{s}-6} ) \big )
\end{equation}
plus 4 times the length of the cuts $t=4$ to $t=4+\frac{32}{\sqrt{s}-6} $ .

For $s> 16$,
\begin{equation}
 L(s) < 4 s .
\end{equation}
We need now a lower bound for $Im f_l (s)$. $Im f_l (s)$ is given by a 
contour integral including the  contribution from the cuts and the ellipse.
We use the fact that $Q_l( )$ is a decreasing function for an argument $>1$.
 We limit arbitrarily the integration on the cuts to 
 $$4+\frac{64}{s-16} < t < 4+\frac{64+P(s)}{s-16} , $$  
 where , 
 \begin{equation}
 P(s) < Const \>; \> 4+\frac{64+P(s)}{s-16} < 4+\frac{32}{\sqrt{s}-6} 
 \end{equation}
which is certainly valid for sufficiently large $s$.
 A lower bound on $Im f_l$ is given by
 \begin{eqnarray}
&&  Im f_l > \frac{1}{\pi k^2} Q_l (1+ \frac{1}{s-4}( 8+\frac{128+2P(s)}{s-16} ) )
\nonumber \\
&& \times  \int _{4+\frac{64}{s-16}  } ^{ 4+\frac{64+P(s)}{s-16}  } \rho (s,t) dt 
\nonumber \\
&&  -  \frac{1}{ 4 \pi k^2} B(s) L(s)  Q_l (1+ \frac{1}{s-4}( 8+\frac{64}{\sqrt{s}-6} ) )                              
 \end{eqnarray}
 Notice that $\rho (s,t)$ according to \cite{Mahoux-Martin} is strictly positive,
given by the double integral of Mandelstam in the strip $4<t< 4+32/(\sqrt{s}-6) $.

Now,given $B(s), L(s)$ and $\rho (s,t) $ it is possible to prove that $|f_l|^2 $ 
is strictly less than $Im f_l $ for $l$ sufficiently large. We have

\begin{equation}
 |f_l|^2< \frac{1}{(4 \pi k^2)^2} Q_l ^2 (1+\frac{8}{s-4} )|B(s)|^2 |L(s)|^2
\end{equation}
and so
\begin{eqnarray}
&& \frac {Im f_l } {|f_l|^2 } >\frac{16 \pi k^2}{|B(s)|^2 |L(s)|^2 }
 \frac{Q_l(x_1)}{Q_l ^2(x_2)} 
\nonumber \\
&&\times \int _{4+\frac{64}{s-16}  } ^{ 4+\frac{64+P(s)}{s-16}  } \rho (s,t) dt 
\nonumber \\
&&-\frac{4 \pi k^2}{B(s)L(s)} \frac{Q_l(x_3)}{Q_l ^2(x_2)},
 \end{eqnarray}
where we define,
\begin{eqnarray}
 x_1&=& 1+ \frac{1}{s-4}( 8+\frac{128+2P(s)}{s-16} )       ,\nonumber \\
 x_2&=&1+\frac{8}{s-4}  ,\nonumber \\
 x_3&=&1+ \frac{1}{s-4}( 8+\frac{64}{\sqrt{s}-6} ).
\end{eqnarray}
It is convenient to denote,
\begin{eqnarray}
 R_1&=&\frac{2x_2^2-1
 + \sqrt {(2x_2^2-1 )^2-1 }} {x_1+\sqrt{x_1^2-1 } } \nonumber \\
 R_2&=&\frac
 {x_3+\sqrt{x_3^2-1 } }  {x_1+\sqrt{x_1^2-1 } } \>.
\end{eqnarray}
Note that , $x_2 < x_1 $, and for $s$ sufficiently large,
\begin{eqnarray}
  x_1<x_3\>,and\>x_1<2x_2^2-1 , \nonumber \\
 hence,\> R_1 > 1; \> R_2 > 1 \>.
\end{eqnarray}
 We now obtain bounds on the relevant Legendre functions. Using the results (\ref{Q_l^2}) and (\ref{ratio}) in the Appendix,
 we have,
\begin{eqnarray}
 \frac{Q_l(x_1)}{Q_l ^2(x_2)} \geq \frac {1} {2x_2 Q_0 (x_2)} \frac{Q_l(x_1)}{Q_l (2 x_2 ^2 -1)} \nonumber \\
  \geq \frac {1} {2x_2 Q_0 (x_2)} R_1 ^{l+1}.
\end{eqnarray}
Further Equations (\ref{ratio}) and (\ref{lb}) in the Appendix yield,
\begin{equation}
  \frac{Q_l(x_3)}{Q_l ^2(x_2)} \leq \sqrt{\frac{2(l+1)}{\pi} } \big( \frac {R_1} {R_2} \big )^{l+1},
\end{equation}
and Eqn (\ref{ratio}) gives,
\begin{equation}
 \frac{Q_l(x_3)}{Q_l (x_1)} \leq \big( \frac {1} {R_2} \big )^{l+1}.
\end{equation}

We now have,
\begin{eqnarray} \label{estimate}
&& \frac {Im f_l } {|f_l|^2 } >\frac{16 \pi k^2}{|B(s)|^2 |L(s)|^2 } \frac {1} {2x_2 Q_0 (x_2)} R_1 ^{l+1}\nonumber \\
&&\times \int _{4+\frac{64}{s-16}  } ^{ 4+\frac{64+P(s)}{s-16}   } \rho (s,t) dt \nonumber \\
&&-\frac{4 \pi k^2}{B(s)L(s)} \sqrt{\frac{2(l+1)}{\pi} } \big( \frac {R_1} {R_2} \big )^{l+1},
\end{eqnarray}
without asymptotic approximations.

 For $ s \rightarrow \infty $,
 \begin{eqnarray}
 && x_1+\sqrt{x_1^2-1} \sim 1 +\frac {4}{\sqrt {s}} +\frac{8}{s}+..,\nonumber \\
&&  x_3+\sqrt{x_3^2-1} \sim 1 +\frac {4}{\sqrt {s}} +\frac{24}{s}+..,\nonumber \\
&& 2x_2^2-1 + \sqrt {(2x_2^2-1 )^2-1 } \sim 1 +\frac {8}{\sqrt {s}} +\frac{32}{s}+..,\nonumber \\
&&   R_1 \sim  1+4/\sqrt {s},\>and \>(1/R_2) \sim 1-16/s .\> 
 \end{eqnarray}
It is clear that since $R_2 > 1$, for $l$ large enough,  i.e. for 

$$ l > \> L_0(s)= Const. s \ln {s} , s \rightarrow \infty ,$$
the contribution of the first term on the right-hand side of Eqn. (\ref{estimate}) involving 
a positive double spectral function is dominant, and that term implies that   

$$ \frac {Im f_l } {|f_l|^2 } \rightarrow \infty , l > \>  Const. s \ln {s}\>.$$ 
 
 Hence the inelastic cross section is dominant and non-zero for $l> L_0(s)$ .
 The fact that $\rho (s,t)$ is different from zero is essential.
 We now evaluate the lower bound on $Im f_l$ , and hence on the inelastic cross section at high energies.
  
  {\bf III. Lower bound on the double spectral function }.

  We must get a lower bound on $\rho (s,t)$.This is 
  relatively easy. We return to the Mandelstam equation 
  (\ref{Mandelstam} ) for $4<t<16$ and restrict ourselves 
  to the Mahoux-Martin domain (\ref{domain1})  
  of positivity of $\rho (s,t)$. To get a lower bound on 
  $\rho (s,t)$ we shall do rather wild majorizations.
  
  1) We reduce the domain of intgration in the $\theta_1, \theta_2$ plane 
  (\ref{thetadomain})to the union of three regions $A,B,C$ (see figure \ref{Regions A,B,C})

\begin{figure*}[!] 
\includegraphics[width=2.0\columnwidth]{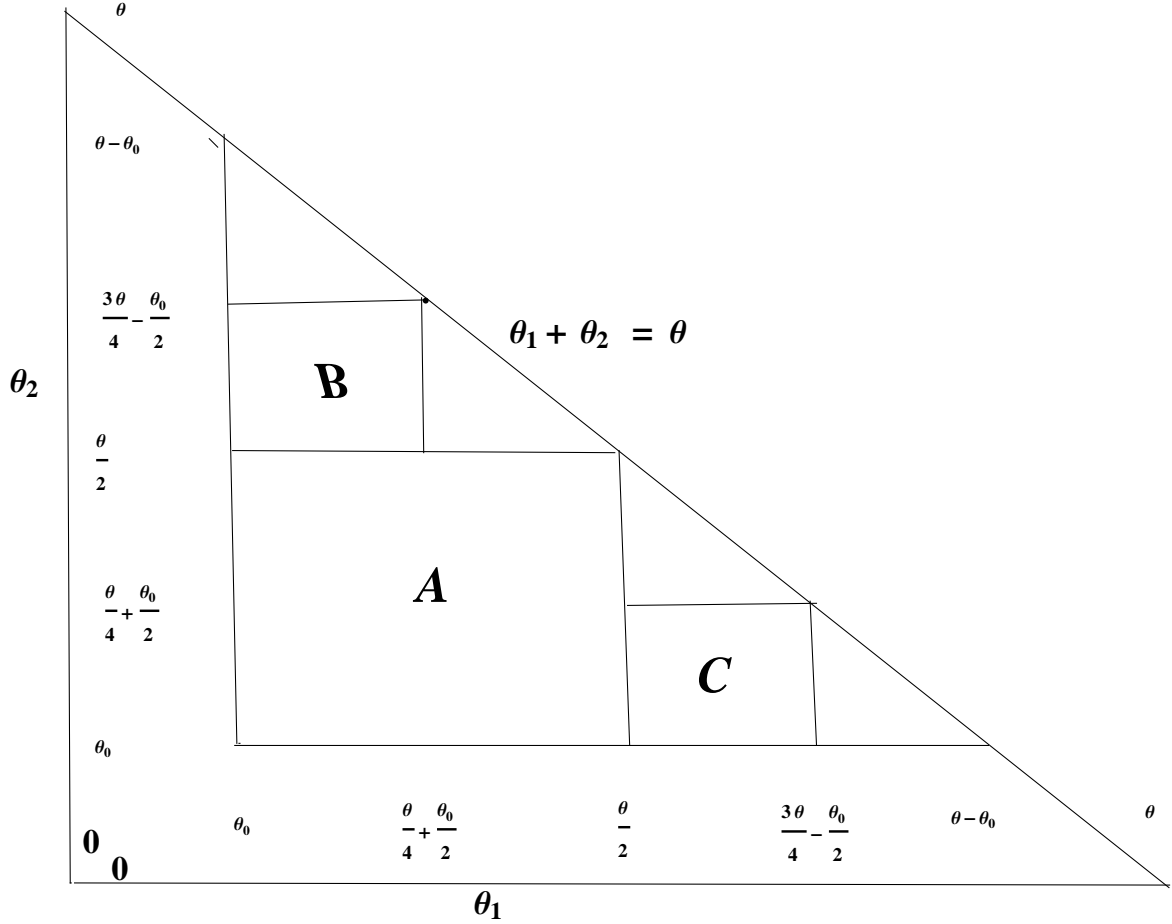}
\caption{
We show the triangular integration region in the $\theta_ 1$, 
  $\theta_ 2 $ plane in Mandelstam's continued elastic unitarity equation in $t-$ channel, defined by $\theta_ 1 \geq \theta_ 0 $, 
  $\theta_ 2 \geq \theta_ 0 $, and $\theta_ 1 + \theta_ 2 \leq \theta $.The sub-regions $A, B, C$ are used to calculate lower bounds on the double 
spectral function .
}
\label{Regions A,B,C}
\end{figure*}
  
  \begin{eqnarray}\label{ABC}
  &&A: \> \> \theta_0 \leq \theta_i \leq \theta_M \equiv \theta /2, \> \> i=1,2 \nonumber \\
 &&i.e.  z_0 \leq z_i \leq z_M \equiv \sqrt{\frac{1+z}{2} } \>\> i=1,2 \nonumber \\ 
 &&B: \theta_0 \leq \theta_1 \leq \theta_{1M} \equiv (\theta /4 +\theta_0/2)\nonumber\\
 &&i.e.  z_0 \leq z_1 \leq z_{1M} \equiv \cosh {(\theta /4 +\theta_0/2) }  \nonumber \\
 && \theta /2 \leq \theta_2 \leq \theta_{2M} \equiv (3 \theta /4 -\theta_0/2)\nonumber \\
 &&i.e.  z_M \leq z_2 \leq z_{2M} \equiv \cosh {(3 \theta /4 -\theta_0/2) }  \nonumber \\
 &&C: \theta_0 \leq \theta_2 \leq \theta_{1M} \nonumber\\
 &&i.e.  z_0 \leq z_2 \leq z_{1M}  \nonumber \\
 && \theta /2 \leq \theta_1 \leq \theta_{2M} \nonumber \\
 &&i.e.  z_M \leq z_1 \leq z_{2M}  
  \end{eqnarray}
  Notice that  under $z_1\leftrightarrow z_2 $ , the regions $B \leftrightarrow C$ and $A\leftrightarrow A $.

  2) Using Eqns. (\ref{H}), (\ref{z+-}), we shall replace $H(z,z_1,z_2)$ in the denominator 
  by simple upper bounds on it in the three regions:
  \begin{eqnarray}
  && A: H(z,z_1,z_2) \leq (z-1)^2 \nonumber \\
  && B,C :  H(z,z_1,z_2) \leq (z-z_-)^2 \leq (z-z_3)^2 \>; \nonumber \\
  && z_3 \equiv \cosh {(\theta /4 -\theta_0/2)}.
  \end{eqnarray}
It will be convenient to define,
\begin{equation}
 \big (z_M,z_{1M},z_{2M},z_3 \big )= 1+ \frac{2}{t-4}\big ( s_M,s_{1M},s_{2M},s_3 \big )
\end{equation}

 3) Since we are in the Mahoux-Martin domain in which  $F_s (s_1,t)$ and  $F_s (s_2,t)$ have 
 convergent partial wave expansions with positive partial waves, and $t$ is positive, 
 the absorptive parts obey the bounds,
 \begin{equation} \label{simple}
  F_s (s_i,t) \geq F_s(s_i,0)=\frac {k_i \sqrt{s_i}}{16 \pi} \sigma_{tot} (s_i); \> i=1,2.
 \end{equation}
They also obey stronger bounds in terms of $\sigma_{tot} (s_i)$ , originally derived by Martin \cite{Martin1966}
or $0<t<4$, but also valid for $4 <t<4+\frac{32}{\sqrt{s}-6}$ under the present assumptions. At high energies 
they have the simple form,
\begin{eqnarray}
&& F_s(s_i,t) \geq F_s(s_i,0) \frac{2 I_1(x_i)} {x_i} (1 + O(1/\sqrt{s_i }));\nonumber \\
&& x_i=\sqrt{t\sigma_{tot }(s_i)/(4 \pi) };\> i=1,2 
\end{eqnarray}

Using the majorizations 1), 2) and the weaker bound (\ref{simple}) in 3), we obtain 

\begin{equation}
 \rho (s,t) \geq \frac {4}{\pi \sqrt {t(t-4)} }\big [\frac {1} {s} I^2 (s_M) +\frac {2} {s-s_3} I (s_{1M}) I (s_{2M}) \big ],
\end{equation}
where the first term in the braces on the right is the contribution of region $A$ and the second term of regions $B$ and $C$ ,

\begin{equation}
 I(s_M)\equiv \int _4 ^{s_M} \frac {ds_1 k_1 \sqrt{s_1} \sigma_{tot} (s_1)}{16\pi},
\end{equation}
and $I (s_{1M})$, $I (s_{2M}) $ are defined similarly by replacing $s_M$ by $s_{1M}$ and $s_{2M}$ respectively.
Note that $s_M$ ,$s_{1M}$ and $s_{2M}$ depend on $s,t$. E.g. 
\begin{equation}
 2 s_M= \sqrt{(t-4)(t-4+s)} -(t-4).
\end{equation}
A simple bound is obtained by retaining only the region $A$. In addition to the above results for general $P(s)$, 
we shall evaluate bounds on $I(s_M), \rho (s,t)$ and the integral over $t$ of $\rho (s,t)$, 
for two simple choices of $P(s)$.

(i) P(s) independent of $s$. Let $P_1< p < P_2$ then we can get a lower bound on the integral over $t$ of $\rho (s,t)$ 
by restricting to the interval
\begin{eqnarray}
&& (64+P_1)/(s-16) < t-4= (64+p)/(s-16) \nonumber \\
&& < (64+P_2)/(s-16). 
\end{eqnarray}
Then, 
$$t(t-4) <  \frac{(64+P_2)(4s +P_2)}{(s-16)^2} $$.
 
For fixed $s$ large enough, $s_M$ is an increasing function of $t$, and hence  
it's minimum value is at the lowest value of $t$,
\begin{eqnarray}
 s_M \geq  (s_M)_{min} \equiv \nonumber \\
 \frac{ \sqrt{(64+P_1)[64+P_1+s(s-16)]}-(64+P_1) }{2(s-16)}
\end{eqnarray}
and 
\begin{equation}
 I(s_M) \geq I((s_M)_{min}.
\end{equation}
Finally we have the bound,
\begin{equation}\label{good bound}
 \int _{4+\frac{64 +P_1}{s-16}  } ^{ 4+\frac{64+P_2}{s-16}   } \rho (s,t) dt 
 \geq \frac{4(P_2-P_1) I^2 ((s_M ){min} ) }{\pi s \sqrt { (64+P_2)(4s+P_2)}   }
\end{equation}
which is positive definite and $> Const. s^{-3/2} $ unless the total cross section vanishes identically 
at all energies upto $ (s_M ){min}$.
 
(ii) $P(s)\rightarrow 0$ for $s\rightarrow \infty$. Then, we integrate over the region,
\begin{eqnarray}
&& 4+\frac{(64 +p_1(s))}{(s-16)} < t = 4+\frac{(64 +p(s))}{(s-16)} \nonumber\\
&& < 4+ \frac{(64 +p_2(s))}{(s-16)} 
\end{eqnarray}
where $p_1(s)$ and $p_2(s) \rightarrow 0,$ for $s \rightarrow \infty$,
and we get $s_M-4 \sim p(s)/32 \rightarrow 0 $. In the integral defining $I(s_M)$ we can 
therefore replace 
\begin{equation}
 \sigma_{tot} \rightarrow 8\pi a_0 ^2 ,
 \end{equation}
where $a_0$ is the $S-$ wave scattering length, and obtain 
\begin{equation}
 I^2(s_M) \rightarrow (p(s)/32)^3 a_0 ^4 /9 \> \geq (p_1(s)/32)^3 a_0 ^4 /9 .
\end{equation}
Finally we obtain for $s\rightarrow \infty$,  $p_1(s)$ and $p_2(s) \rightarrow 0,$ as slowly 
as we like,
 \begin{equation}
 \int _{4+\frac{64 +p_1(s)}{s-16}  } ^{ 4+\frac{64+p_2(s)}{s-16}   } \rho (s,t) dt 
 \geq \frac{p_2(s)-p_1(s)}{36\pi s^{3/2} } (\frac{p_1(s)}{32})^3 a_0 ^4 .
\end{equation}
 This bound is of interest as it shows that the asymptotic inelastic cross section cannot vanish  
 if the $S-$wave scattering length is non-zero. However,
 the bound (\ref{good bound}) is preferable as it does not need any asymptotic approximation.

  {\bf IV. Asymptotic behaviour of the lower bound on inelastic cross section
 and discussion of the assumptions}.
 
 Now we know that ,above a certain energy, the inelastic cross 
 section cannot be zero.A lower bound can be obtained if we 
 know something about $B(s)$ and if we accept the postulated 
 analyticity.If we believe in the validity of the Mandelstam 
 representation with a finite number of subtractions, $B(s)=s^N$.
 In fact we tend to believe that $B(s)=s^2/s_0^2$ , because we 
 postulate an ellipse (with cuts) which in the limit of 
 high energy coincides with the ellipse with foci $t=0, u=0$ 
 and extremities $t=4, u=4$.Inside this ellipse the absorptive 
 part $F_s(s,t)$ is maximum for $t$ real, $0<t<4$, and the integral
 \begin{equation}
  \int \frac {F_s(s,t) ds } {s^3 } < \infty
 \end{equation}
which means that $F_s(s,t)$ is almost everywhere less than $s^2$.
  Concerning the dispersive part which is, modulo subtractions,
  the Hilbert transform of the absorptive part we have a rather 
  tricky argument to show again that it is almost everywhere bounded 
  by $s^{2+\epsilon}$, $\epsilon$ arbitrarily small , for any $t$ for which dispersion relations 
  are valid.But we shall not use that result here.
  
 Using the lower bound on the integral of the double spectral function, and 
 $B(s)=s^N$,we deduce that the ratio of the contributions of the cut term and 
 the elliptical contour ($\Gamma$) term  to $Im f_l$ 
 goes to infinity if 
 \begin{equation}
  l > L_0(s) = \frac{(N+5/2)}{16} s \ln {s} .
 \end{equation}
The ratio of the  contribution of the cut term  to $Im f_l$  to the upper bound on $|f_l|^2 $
 goes to infinity for a much smaller value, viz. if 
  \begin{equation}
  l > L_1(s) = \frac{\sqrt{s}}{4} (2N+5/2) \ln {s} .
 \end{equation}
Hence, summing the contributions of partial waves with $l > L_0(s)$ we see that for $ s\rightarrow \infty$,
\begin{equation}
 \sigma_{inel} (s) > \frac{Const}{s^{5/2} }\exp {[-\frac{\sqrt{s}}{4} (N+5/2) \ln {s} ]}.
\end{equation}

 {\bf V. Real Pions of Isotopic Spin 1}.
 Let $F^{(I)}(s,t,u)$ denote the $\pi \pi\rightarrow \pi \pi$ amplitudes with total 
 iso-spin $I$ in the $s$-channel, $I=0,1,2$, and  $F^{(I)}(t,s,u)$ the amplitudes with 
 iso-spin $I$ in the $t$-channel. They are related by the crossing matrix $C_{st}$,
 \begin{eqnarray} \label{crossing}
 \begin{bmatrix}
  F^{(0)}(t,s,u)\\
  F^{(1)}(t,s,u)\\
  F^{(2)}(t,s,u)
  \end{bmatrix}
  =C_{st}
  \begin{bmatrix}
  F^{(0)}(s,t,u)\\
  F^{(1)}(s,t,u)\\
  F^{(2)}(s,t,u)
  \end{bmatrix},\nonumber\\
  C_{st}=
  \begin{bmatrix}
  1/3 & 1 & 5/3 \\
  1/3 & 1/2 & -5/6\\
  1/3 & -1/2 & 1/6
  \end{bmatrix}.
 \end{eqnarray}
We do {\bf not} assume the unsubtracted Mandelstam representation,
\begin{eqnarray}
 F^{(I)}(s,t,u)= \frac{1}{\pi ^2} \int \int \frac {\rho_{st}^{(I)}(s',t')ds' dt'}{(s'-s)(t'-t)} + \nonumber\\
 \frac {1}{\pi ^2} \int \int \frac {\rho_{su}^{(I)}(s',u')ds' du'}{(s'-s)(u'-u)} +\frac{1}{\pi ^2} \int \int \frac {\rho_{tu}^{(I)}(t',u')dt' du'}{(t'-t)(u'-u)}\nonumber \\.
\end{eqnarray}
However, we use the definitions
\begin{eqnarray}
 F_{st}^{(I)}(s,t,u)=\rho_{st}^{(I)}(s,t), F_{su}^{(I)}(s,t,u)=\rho_{su}^{(I)}(s,u),\nonumber\\
 F_{tu}^{(I)}(s,t,u)=\rho_{tu}^{(I)}(t,u),
\end{eqnarray}
and Eq. (\ref{crossing}) then implies that
\begin{equation}
 F_{st}^{(I)}(t,s,u)=\rho_{st}^{(I)}(t,s)=\sum _{I'=0,1,2} C_{st}^{II'}\rho_{st}^{(I')}(s,t).
\end{equation}
Note that in  $\rho_{st}^{(I)}(t,s)$ and $\rho_{st}^{(I')}(s,t) $, the superscripts $I,I'$ denote iso-spins in the channel specified 
by the first argument, viz. t-channel and s-channel respectively.
The Mandelstam unitarity equations for $t$-channel Iso-spin $I$, and $4\leq t \leq16$ is given by Mahoux and Martin \cite{Mahoux-Martin},
\begin{eqnarray} \label{Mahoux}
&&  \rho ^{(I)} (t,s) = \frac{2 \rho (t)}{\pi} \int \int \frac { dz_{1} dz_{2} \theta (z-z_+)}
   {\sqrt {H(z,z_{1},z_{2})} } G ^{(I)} (t,s_1,s_2),\nonumber \\
&& G ^{(I)} (t,s_1,s_2)= (-1)^I F_s ^{(I)}(t,s_1)F_s ^{(I)*} (t,s_2).
\end{eqnarray}
Crossing, Eq. (\ref{crossing}) ,immediately yields
\begin{eqnarray} \label {Roy}
 && G ^{(I)} (t,s_1,s_2)= \sum_{I',I''=0,1,2}\zeta ^I _{I'I''} F_s ^{(I')}(s_1,t)F_s ^{(I'')}(s_2,t)^* ,\nonumber\\
 && \zeta ^I _{I'I''}=(-1)^I  C_{st}^{II'} C_{st}^{II''},
\end{eqnarray}
where,
\begin{eqnarray}
&& \zeta ^0=
 \begin{bmatrix}
  1/9 & 1/3 & 5/9\\
  1/3 & 1 & 5/3   \\
  5/9 & 5/3 & 25/9 
 \end{bmatrix}
 , \zeta ^1=
 \begin{bmatrix}
  -1/9 & -1/6 & 5/18 \\
  -1/6 & -1/4 & 5/12 \\
  5/18 & 5/12 & -25/36 
 \end{bmatrix} ,\nonumber \\
&&  \zeta ^2 =
 \begin{bmatrix}
  1/9 & -1/6 & 1/18 \\
  -1/6 & 1/4 & -1/12 \\
  1/18 & -1/12 & 1/36 
 \end{bmatrix}
\end{eqnarray}

which are identical to the values  obtained in \cite{Mahoux-Martin}, and quoted again 
for ready reference. We now have, 
\begin{eqnarray} \label{rho_positivity}
&& \rho ^{(I)} (t,s) =  \frac{2 \rho (t)}{\pi} \int \int \frac { dz_{1} dz_{2} \theta (z-z_+)}
   {\sqrt {H(z,z_{1},z_{2})} } \nonumber\\
&& \times \sum_{I',I''=0,1,2}\zeta ^I _{I'I''}   F_s ^{(I')}(s_1,t)F_s ^{(I'')}(s_2,t)^* .
\end{eqnarray}

Mahoux and Martin \cite{Mahoux-Martin} have noted that all the matrix elements of 
\begin{equation}
 \zeta ^0, \zeta ^0 -\zeta ^2, \zeta ^0 +\zeta ^1,\zeta ^0 -2 \zeta ^1, \>and \> \zeta ^0 + 2 \zeta ^2
\end{equation}
are positive, and for $s,t$ in the Mahoux-Martin domain (\ref{domain1}) the $F_s ^{(I)}(s_i,t), i=1,2 $ are positive for the 
relevant values of $s_i$ due to unitarity. From Eqn. (\ref{rho_positivity}),it follows that,
\begin{equation}
 \sum_I \beta_I \zeta^I_{I',I''} >0 ,\>for\>all>I',I'' \Longrightarrow \>\sum_I \beta_I \rho ^{(I)} (t,s) >0.
\end{equation}
Hence,  
\begin{eqnarray}\label{rho_t combinations}
 \rho ^{(0)} (t,s),\rho ^{(0)} (t,s)-\rho ^{(2)} (t,s),\rho ^{(0)} (t,s)+\rho ^{(1)} (t,s),\nonumber\\
 \rho ^{(0)} (t,s)-2\rho ^{(1)} (t,s), \>and\> \rho ^{(0)} (t,s) +2 \rho ^{(2)} (t,s)
\end{eqnarray}
are positive in the Mahoux-Martin domain. We can exploit these results to get bounds on inelastic cross sections 
for real pions (of iso-spin 1).

{\bf New results}. The truncated Froissart-Gribov formula will enable us to obtain lower 
bounds on imaginary parts of $s$-channel partial waves of the following five amplitudes:
\begin{eqnarray}\label{amplitudes}
&& \big (\frac{1}{3} F^{(0)} +F^{(1)}+\frac{5}{3}F^{(2)} \big )(s,t)=F^{(0)}(t,s); \nonumber \\
&&\frac{3}{2} \big (F^{(1)} +F^{(2)}\big )(s,t)= \big ( F^{(0)} -F^{(2)} \big ) (t,s);\nonumber\\
&& \big (\frac {2}{3} F^{(0)} +\frac{3}{2} F^{(1)}+\frac{5}{6} F^{(2)} \big )(s,t)=\big ( F^{(0)}+F^{(1)} \big )(t,s);\nonumber \\
&&\big (-\frac{1}{3} F^{(0)} +\frac{10}{3}F^{(2)} \big ) (s,t)=\big ( F^{(0)} -2F^{(1)} \big )(t,s);\nonumber \\
&& \frac{1}{3} \big (F^{(0)} +2 F^{(2)}\big )(s,t)=  \frac{1}{3} \big (F^{(0)} +2 F^{(2)}\big )(t,s),
\end{eqnarray}
where the right-hand sides correspond to the t-channel Iso-spin combinations in Eq. (\ref{rho_t combinations} ) and 
the left-hand sides are the corresponding linear combinations of s-channel Iso-spin amplitudes.
  These equations are of the form,
  \begin{equation}
     \sum _I \alpha_I F^{(I)}(s,t)= \sum _I \beta_I F^{(I)}(t,s),
   \end{equation}
 where the coefficients $\alpha_I$ and $\beta_I$ can be read off the Equations (\ref{amplitudes}) .E.g. $\alpha_0=\beta_0=1/3, \alpha_2=\beta_2=2/3,
 \alpha_1=\beta_1=0$ 
 for the last amplitude which is just the $\pi ^0 \pi ^0 \rightarrow  \pi ^0 \pi ^0 $ amplitude,
\begin{equation}
 F^{00} \equiv \frac{1}{3} \big (F^{(0)} +2 F^{(2)}\big ).
\end{equation}
The partial waves given by the truncated Froissart-Gribov formula are then, 
for even $l+I$,
 \begin{eqnarray}
 &&  \sum_I \alpha_I f_l^I(s)  \nonumber \\
 && = \frac{1}{4 i\pi k^2} \int _\Gamma 
   Q_l(1+\frac{2t}{s-4}) \sum _I \beta_I F^{(I)}(t,s) dt +\nonumber \\
 &&  \frac{1}{\pi k^2} \int _4 ^{4+\frac{32}{\sqrt{s}-6}} 
   Q_l(1+\frac{2t}{s-4}) \sum _I \beta_I F_t^{(I)}(t,s) dt 
  \end{eqnarray}
  
and
 \begin{eqnarray}
 &&  \sum_I \alpha_I Im f_l^I(s) \nonumber \\
 && =\frac{1}{4 i\pi k^2} \int _\Gamma 
   Q_l(1+\frac{2t}{s-4}) \sum _I \beta_I F_s^{(I)}(t,s) dt +\nonumber \\
&&  \frac{1}{\pi k^2} \int _{4+\frac{64}{s-4}} ^{4+\frac{32}{\sqrt{s}-6}} 
   Q_l(1+\frac{2t}{s-4}) \sum _I \beta_I \rho ^{(I)} (t,s) dt 
  \end{eqnarray}
  As before, $\Gamma$ is an ellipse with foci at $t=0$ and $u=0$, and right 
extremity at $t=4+\frac{32}{\sqrt{s}-6}$. 
As for pions without iso-spin, we prove, if we only use the region $A$ in Fig. (\ref{Regions A,B,C}) ,that the combinations $\sum _I \beta_I \rho ^{(I)} (t,s) $ 
on the right-hand side are not ony positive, but also have a lower bound,
\begin{equation}
 \sum_I \beta_I \rho ^{(I)} (t,s) \geq  \frac {4}{\pi s\sqrt {t(t-4)} } \sum_I \beta_I \zeta^I_{I',I''} I^{I'}(s_M) I^{I''}(s_M) 
\end{equation}
provided that $\sum_I \beta_I \zeta^I_{I',I''} >0$ , for all $I',I'' $, and
\begin{equation}
 I^{I'}(s_M)\equiv \int _4 ^{s_M} \frac {ds_1 k_1 \sqrt{s_1} \sigma_{tot}^{(I')} (s_1)}{16\pi}.
\end{equation}

We can now obtain lower bounds on the cut contributions to linear combinations of imaginary parts of $s$-channel partial waves ,
\begin{eqnarray}
&&  1/3 f_l^{(0)} +f_l^{(1)}+5/3 f_l^{(2)} ,3/2(f_l^{(1)}+ f_l^{(2)}),\nonumber\\
&&  2/3f_l^{(0)} +3/2 f_l^{(1)}+5/6 f_l^{(2)},-1/3 f_l^{(0)} +10/3 f_l^{(2)},\nonumber\\
&& 1/3 f_l^{(0)} +2/3 f_l^{(2)} 
\end{eqnarray}
from lower bounds respectively on the combinations of $\rho ^{(I)}(t,s)$ given in Eqn. (\ref{rho_t combinations}).
The contributions to these imaginary parts from the elliptical contours $\Gamma$ are negligible for 
$l>L_0(s)$;the elastic pion-pion cross sections (including also $\pi ^0 \pi ^0 \rightarrow  \pi ^+ \pi ^- $ 
cross sections are negligible for $l>L_1(s)$, and hence also for $l>L_0(s)$. On summing the contributions of 
$l>L_0(s)$ lower bounds on $Im [3/2(f_l^{(1)}+ f_l^{(2)})]$, 
and $Im [1/3 f_l^{(0)} +2/3 f_l^{(2)}]$ to inelastic cross sections we obtain  the three inequalities,

\begin{eqnarray}
  \sigma_{inel}^{(1)} (s),\sigma_{inel}^{(2)} (s), \sigma_{inel}^{\pi ^0 \pi ^0 } (s) \nonumber\\
  > \frac{Const}{s^{5/2} }\exp {[-\frac{\sqrt{s}}{4} (N+5/2) \ln {s} ]}.
\end{eqnarray}
 
 {\bf APPENDIX. Bounds on Associated Legendre Functions }.
 
 We derive bounds on $Q_l(x)$ for real $l$ and complex $x$ using the integral representation,
 
 \begin{equation}
  Q_l (x)=\int _0 ^\infty \frac {dt}{(x+ \sqrt{x^2-1}\cosh {t}  )^{l+1} }
 \end{equation}

 1.{\bf Upper Bound}.  For $x$ real $>1$ , 
 \begin{equation} Q_l (x) \le (x+ \sqrt{x^2-1} )^{-l} Q_0 (x) \>.
 \end{equation}
 This is obvious because , $x+ \sqrt{x^2-1}\cosh {t} \ge(x+ \sqrt{x^2-1}) $.
 
 2.  {\bf Lower Bound}.  For $x$ real $>1$ ,   
 \begin{equation} \label{lb}
 Q_l (x) \ge (x+ \sqrt{x^2-1} )^{-l-1} \sqrt{\frac{\pi}{2(l+1)} } \>.
 \end{equation}
 Proof. It is obvious that 
 \begin{equation} \label{lb1}
  Q_l (x) \ge (x+ \sqrt{x^2-1} )^{-l-1} \int _0 ^\infty \frac {dt}{(\cosh {t}  )^{l+1} },
 \end{equation}
 because $(x+ \sqrt{x^2-1}\cosh {t}) \le (x+ \sqrt{x^2-1}) \cosh {t} $. The integral 
 on the right-hand side is exactly known \cite{Gradshteyn},
 \begin{equation}
  \int _0 ^\infty \frac {dt}{(\cosh {t}  )^{l+1} }=\frac{2^{l-1}}{\Gamma (l+1)} \Gamma ^2  (\frac{l+1}{2}),
 \end{equation}
but we shall need only a lower bound on it.Using $\cosh {t} \le \exp {(t^2/2) }$ we have,
\begin{eqnarray}
 \int _0 ^\infty \frac {dt}{(\cosh {t}  )^{l+1} } \ge \int _0 ^\infty dt \exp{(-t^2 (l+1)/2 ) } \nonumber \\
 =\sqrt{\frac{\pi}{2(l+1)} } \>.
\end{eqnarray}
Inserting this in Eq. (\ref{lb1} ) we obtain the quoted lower bound Eq. (\ref{lb} ).
 
 3. {\bf Upper bound on an ellipse in complex z-plane}. We prove that for 
 real values of $\theta_1,\theta_2 $,
  
  \begin{equation}
   \big |Q_l (\cosh {(\theta _1 +i\theta _2) } ) \big | \leq  Q_l (\cosh {\theta _1 } ), 
  \end{equation}
i.e. geometrically, for $z$ on an ellipse with foci $-1$ and $1$, and right extremity $z_0= \cosh {\theta_1 }$,

\begin{equation}
 \big | Q_l (z) \big | \le  Q_l (z_0) \> for \> z=\cosh {(\theta _1 +i\theta _2) }.
\end{equation}
The denominator in the integral representation of $Q_l(z)$ is $|D(z,t)|^{l+1}$, where 
\begin{equation}
 D(z,t)=\cosh {(\theta _1 +i\theta _2) } +\cosh {t} \sinh {(\theta _1 +i\theta _2) }.
\end{equation}
 It suffices to prove that    
 \begin{equation} \label{Dz}
  |D(z,t)| > D(z,t)|_{\theta_2 =0 }.
 \end{equation}
Trigonometric identities yield,
\begin{eqnarray}
 |D(z,t)|^2 &=& D(z,t) D(z,t)^* =\frac{1}{2} \cosh {2\theta_1} (1+ \cosh ^2 {t} ) \nonumber \\
 &+&\cosh {t} \sinh {2\theta_1} -\frac{1}{2} \cos {2\theta_2} \sinh ^2 {t}.
\end{eqnarray}
Minimising over $\theta_2 $ now yields the desired result, Eq.(\ref{Dz}).

4. {\bf Upper bound on $Q_l (x)$ in terms of $Q_0 (x)$ and $Q_l(2x^2-1)$ for $x>1$ }. We prove that,

\begin{equation} \label{Q_l^2}
 Q_l^2(x) \le 2x Q_0(x) Q_l (2x^2-1)\>, for\> x >1\>.
\end{equation}
(i) The integral representation of $Q_l(x)$ and Schwarz inequality yield,
\begin{equation}
 Q_l^2(x) \le  Q_0(x) Q_{2l} (x)\>.
\end{equation}
Hence, to prove (\ref{Q_l^2}) it will be sufficient to prove that 
\begin{equation}
 Q_{2l}(x) \le 2x Q_l (2x^2-1).
\end{equation}
Using,
  \begin{eqnarray}
& &  (x+ \sqrt{x^2-1}\cosh {t}  )^2 =2x^2-1  \nonumber\\ 
 & &+\sqrt{(2x^2-1)^2-1}\cosh {t} + (x^2-1) \sinh ^2{t} \nonumber\\
& &  \ge 2x^2-1+ \sqrt{(2x^2-1)^2-1}\cosh {t},
 \end{eqnarray}
 and 
 \begin{eqnarray}
  2x^2-1+ \sqrt{(2x^2-1)^2-1}\cosh {t}\nonumber\\
  =2x  (x+ \sqrt{x^2-1}\cosh {t}  ) -1\>,
 \end{eqnarray}
we have the required result
\begin{eqnarray}
 Q_{2l}(x) \le \int _0 ^{\infty} dt \> (2x- \frac{1}{x+ \sqrt{x^2-1}\cosh {t} } ) \nonumber \\
 \times \frac{1}{ (2x^2-1+ \sqrt{(2x^2-1)^2-1}\cosh {t} )^{l+1}} \nonumber \\
 \le 2x Q_l (2x^2-1)\>.
 \end{eqnarray}

  5. {\bf Upper bound on $Q_l (x)/Q_l(z) $ for $x> z >1$ }.
 We prove that for $x> z >1$ 
 \begin{equation} \label{ratio}
  \frac{Q_l(x)}{Q_l(z)} \le \bigg ( \frac{z+ \sqrt{z^2-1} }{x+ \sqrt{x^2-1} }\bigg )^{l+1} \le \bigg ( \frac{1+ \sqrt{2(z-1)} }{1+ \sqrt{2(x-1)} }\bigg )^{l+1} . 
 \end{equation}
Using the integral representation we obtain,
\begin{eqnarray}
\frac {d}{dz} \bigg ( ( z+ \sqrt{z^2-1} )^{l+1} Q_l(z) \bigg )= -\frac{l+1 } {\sqrt {z^2-1} } \nonumber \\
 \times ( z+ \sqrt{z^2-1} )^l  \int _0 ^\infty \frac {dt ( \cosh{t} -1)}{(z+ \sqrt{z^2-1}\cosh {t}  )^{l+2} } \nonumber \\
\le 0 \>,
\end{eqnarray}
which implies the left-hand side of the inequality (\ref{ratio}). The right-hand side now follows if,
\begin{equation}
 \bigg ( \frac{z+ \sqrt{z^2-1} }{x+ \sqrt{x^2-1} }\bigg ) \le \bigg ( \frac{1+ \sqrt{2(z-1)} }{1+ \sqrt{2(x-1)} }\bigg ),
 \end{equation}
 or if,
 \begin{equation}
  \bigg ( \frac{z+ \sqrt{z^2-1} }{1+ \sqrt{2(z-1)} }\bigg ) \le \bigg ( \frac{x+ \sqrt{x^2-1}} {1+ \sqrt{2(x-1)} }\bigg ),
 \end{equation}
for $x> z >1$. This holds since the left-hand side of the above inequality is an increasing function of $z$ for $z > 1$. 
 
  {\bf Acknowledgements}.
  One of us (AM) is grateful to Jo$\tilde{a}$o Penedones for inviting him to participate in a workshop on 
  S-matrix bootstrap in Lausanne, and to Miguel F. Paulos for raising the question whether inelasticity 
  could be arbitrarily small ;(SMR) wishes to thank the Indian National Science Academy for the 
  INSA Honorary Scientist position.

\end{document}